\documentclass[aps,prb,twocolumn,amsmath,amssymb,groupedaddress,longbibliography
]{revtex4-2}

\usepackage{graphicx}
\usepackage{dcolumn}
\usepackage{bm}
\usepackage{multirow}
\usepackage{braket}
\usepackage{color}
\usepackage{ulem}
\usepackage{times}
\usepackage{booktabs}

\begin{document}
\title{
Square skyrmion crystal in centrosymmetric itinerant magnets
}
\author{Satoru~Hayami and Yukitoshi~Motome}
\affiliation{Department of Applied Physics, the University of Tokyo, Tokyo 113-8656, Japan}

\begin{abstract}
We theoretically investigate the origin of the square-type skyrmion crystal in centrosymmetric itinerant magnets, motivated from the recent experimental finding in GdRu$_2$Si$_2$ [N. D. Khanh {\it et al.}, Nat. Nanotech. {\bf 15}, 444 (2020)].  
By simulated annealing for an effective spin model derived from the Kondo lattice model on a square lattice, we find that a square skyrmion crystal composed of a superposition of two spin helices is stabilized in a magnetic field by synergy between the positive biquadratic, bond-dependent anisotropic, and easy-axis anisotropic interactions. 
This is in stark contrast to triangular skyrmion crystals which are stabilized by only one of the three, suggesting that the square skyrmion crystal is characteristic of itinerant magnets with magnetic anisotropy. 
We also show that a variety of noncollinear and noncoplanar spin textures appear depending on the model parameters as well as the applied magnetic field. 
The present systematic study will be useful not only for identifying the key ingredients in GdRu$_2$Si$_2$ but also for exploring further skyrmion-hosting materials in centrosymmetric itinerant magnets.
\end{abstract}

\maketitle

\section{Introduction}
A magnetic skyrmion has attracted great interest owing to rich physics emerging from its topological spin texture~\cite{Bogdanov89,Bogdanov94,rossler2006spontaneous,Muhlbauer_2009skyrmion,yu2010real,nagaosa2013topological,fert2017magnetic}. 
For example, a periodic arrangement of the skyrmions, which is referred to as the skyrmion crystal (SkX), gives rise to a giant topological Hall effect~\cite{Bruno_PhysRevLett.93.096806,Neubauer_PhysRevLett.102.186602,Kanazawa_PhysRevLett.106.156603}, Nernst effect~\cite{Shiomi_PhysRevB.88.064409,mizuta2016large}, and nonreciprocal transport~\cite{Hamamoto_PhysRevB.95.224430,seki2020propagation} through the spin Berry phase mechanism~\cite{berry1984quantal,Loss_PhysRevB.45.13544,Xiao_RevModPhys.82.1959}. 
The topological robustness and unconventional transport properties may provide potential applications to next-generation magnetic memory and logic computing devices in spintronics~\cite{fert2013skyrmions,romming2013writing,fert2017magnetic}. 
While many materials have been found to host the skyrmions thus far, they have been mostly limited to the materials with noncentrosymmetric lattice structures and strong spin-orbit coupling. 
In fact, the SkXs were observed in chiral and polar magnets~\cite{Yi_PhysRevB.80.054416,Mochizuki_PhysRevLett.108.017601,Munzer_PhysRevB.81.041203,seki2012observation,kezsmarki_neel-type_2015,lee2016synthesizing,woo2016observation,soumyanarayanan2017tunable,Takagi_PhysRevLett.120.037203,Sen_PhysRevB.99.134404} where the spin-orbit coupling generates the Dzyaloshinskii-Moriya interaction~\cite{dzyaloshinsky1958thermodynamic,moriya1960anisotropic}. 

Recently, several SkXs exhibiting the giant topological Hall and Nernst effects were discovered also in centrosymmetric $f$-electron compounds, such as triangular-type SkXs in Gd$_2$PdSi$_3$~\cite{kurumaji2019skyrmion,hirschberger2019topological,nomoto2020formation} and Gd$_3$Ru$_4$Al$_{12}$~\cite{hirschberger2019skyrmion}, and a square-type SkX in GdRu$_2$Si$_2$~\cite{khanh2020nanometric,Yasui2020}. 
Due to the centrosymmetric lattice structures, their origin might be attributed to magnetic frustration~\cite{Okubo_PhysRevLett.108.017206,leonov2015multiply,Lin_PhysRevB.93.064430,Hayami_PhysRevB.93.184413,batista2016frustration} or 
effective magnetic interactions arising from the spin-charge coupling between conduction and localized electrons~\cite{Martin_PhysRevLett.101.156402,Akagi_PhysRevLett.108.096401,Hayami_PhysRevB.90.060402,Hayami_PhysRevB.95.224424,Ozawa_PhysRevLett.118.147205} rather than the DM interaction.
In particular, GdRu$_2$Si$_2$ can be a prototype for the SkX originating from the spin-charge coupling, since the crystal structure is tetragonal that is free from geometrical frustration. 
Although the origin was speculated to be four-spin interactions mediated by itinerant electrons in the presence of easy-axis anisotropy~\cite{khanh2020nanometric}, it has not been fully clarified yet from the microscopic point of view. 

In the present study, we theoretically examine an instability toward the square SkX on a centrosymmetric tetragonal lattice in itinerant magnets. 
By performing simulated annealing for an effective spin model which incorporates the itinerant nature of electrons, we show that the square SkX is stabilized by the interplay among the four-spin biquadratic interaction, bond-dependent anisotropic interaction, and easy-axis anisotropic interaction in a magnetic field. 
The SkX is a double-$Q$ state composed of a superposition of two spin helices, similar to the one observed in GdRu$_2$Si$_2$~\cite{khanh2020nanometric}. 
We find that the SkX exhibits a larger scalar spin chirality, which leads to a stronger topological Hall response, for a larger biquadratic interaction and smaller bond-dependent anisotropy. 
Besides the square SkX, we find several noncollinear and noncoplanar spin states depending on the model parameters.  
In particular, different types of double-$Q$ states, which appear next to the square SkX upon increasing or decreasing the magnetic field, well explain the experimental results in GdRu$_2$Si$_2$~\cite{khanh2020nanometric,Yasui2020}. 
We also discuss the stability of the square SkX in comparison with that of triangular SkXs; the interplay among the biquadratic, bond-dependent, and easy-axis anisotropic interactions plays an important role in the square SkX, whereas only one of them can stabilize the triangular ones. 
Our systematic analyses would be a reference to further exploration of skyrmion-hosting materials in centrosymmetric itinerant magnets.

The rest of the paper is organized as follows. 
In Sec.~\ref{sec:Model and method}, we introduce the effective spin model with the biquadratic and anisotropic interactions, and the numerical method to investigate the ground state. 
We discuss the magnetic phase diagram at zero field in Sec.~\ref{sec:Zero-field phase diagram}. 
In Sec.~\ref{sec:Skyrmion crystal in a finite field}, we show the results in a magnetic field and identify the key ingredients for the square SkX. 
We discuss the results in comparison with the experiments for GdRu$_2$Si$_2$ in Sec.~\ref{sec:Discussion}. 
We also compare the stability of the square SkX with the triangular one. 
Section~\ref{sec:Summary} is devoted to the summary. 
In Appendix~\ref{sec:appendix}, we show the effect of the magnetic field on the double-$Q$ state which is not focused on in the main text.

\section{Model and method}
\label{sec:Model and method}

We consider an effective spin model on the basis of the Kondo lattice model consisting of itinerant electrons and localized spins~\cite{Hayami_PhysRevB.95.224424,Hayami_PhysRevLett.121.137202,hayami2020multiple,Su_PhysRevResearch.2.013160,Yasui2020}, whose Hamiltonian is given by  
\begin{align}
\label{eq:Model}
\mathcal{H}=  &2\sum_{\bm{q}}
\left( -J  \lambda_{\bm{q}} 
+\frac{K}{N} \lambda_{\bm{q}} ^2 \right)  -H \sum_i S_i^z,
\end{align}
with 
\begin{align}
\label{eq:Gamma}
\lambda_{\bm{q}}= \sum_{\alpha \beta}\Gamma^{\alpha\beta}_{\bm{q}} S^\alpha_{\bm{q}} S^\beta_{-\bm{q}}, 
\end{align}
where the localized spins $\bm{S}_i$ at site $i$ form a square lattice with the number of spins $N$. 
We regard $\bm{S}_i$ as a classical spin with a fixed length $|\bm{S}_i|=1$ for simplicity. 
$\bm{S}_{\bm{q}}$ is the Fourier transform of $\bm{S}_i$. 
The first term in Eq.~(\ref{eq:Model}) consists of the bilinear and biquadratic interactions in momentum ($\bm{q}$) space, whose coupling constants are represented by $J$ and $K$, respectively. 
$\Gamma^{\alpha\beta}_{\bm{q}}$ in Eq.~(\ref{eq:Gamma}) is a $\bm{q}$-dependent dimensionless form factor to represent the magnetic anisotropy that satisfies the fourfold rotational symmetry of the square lattice~\cite{Yasui2020}. 
The second term in Eq.~(\ref{eq:Model}) represents the Zeeman coupling to an external magnetic field $H$ along the $z$ direction. 

The effective spin model with the momentum-space interactions is obtained from the Kondo lattice model by using the perturbation expansion in terms of the spin-charge coupling between itinerant electrons and localized spins~\cite{Akagi_PhysRevLett.108.096401,Hayami_PhysRevB.90.060402,Hayami_PhysRevB.94.024424,Hayami_PhysRevB.95.224424}. 
The bilinear term is derived from the lowest-order expansion, which is referred to as the Ruderman-Kittel-Kasuya-Yosida (RKKY) interaction~\cite{Ruderman,Kasuya,Yosida1957}. 
Meanwhile, the biquadratic term is one of the second lowest-order contributions, which plays a crucial role in stabilizing noncoplanar spin textures composed of superpositions of multiple helices~\cite{Hayami_PhysRevB.95.224424}. 
The coupling constants $J$ and $K$ depend on the electronic state of the itinerant electrons, such as the band filling and hopping parameters. 
We take $J=1$ as an energy unit and $K>0$. 

In order to investigate the magnetic phase diagram in the model in Eq.~(\ref{eq:Model}), we simplify the interaction term by focusing on the situation where the magnetic bare susceptibility of the itinerant electrons shows maxima at $\bm{Q}_1=(Q,0)$ and $\bm{Q}_2=(0,Q)$, which are compatible with the fourfold rotational symmetry. 
We take $Q=\pi/3$ without loss of generality. 
In other words, we ignore the contributions from 
the interactions except for $\bm{Q}_1$ and $\bm{Q}_2$. 
Then, only the form factors at $\bm{Q}_1$ and $\bm{Q}_2$, $\Gamma_{\bm{Q}_1}$ and $\Gamma_{\bm{Q}_2}$, are taken into account, which are given by 
\begin{align}
\label{eq:Gamma1}
\Gamma_{\bm{Q}_1}&=\left(
\begin{array}{ccc}
\Gamma^{\rm iso}-I^{\rm BA} & 0 & 0\\
0 & \Gamma^{\rm iso}+I^{\rm BA} &0 \\
0 & 0 & \Gamma^{\rm iso}+I^{z}
\end{array}
\right), \nonumber \\
\Gamma_{\bm{Q}_2}&=\left(
\begin{array}{ccc}
\Gamma^{\rm iso}+I^{\rm BA} & 0 & 0\\
0 & \Gamma^{\rm iso}-I^{\rm BA} &0 \\
0 & 0 & \Gamma^{\rm iso}+I^{z}
\end{array}
\right). 
\end{align}
Here, $\Gamma^{\rm iso}$ represents the isotropic form factor; 
we take $\Gamma^{\rm iso}=1$. 
Meanwhile, $I^{\rm BA}$ and $I^z$ represent the anisotropic form factors which are taken to be invariant under the fourfold rotational operation. 
These anisotropic interactions arise from the spin-orbit coupling under the crystalline electric field~\cite{khomskii2003orbital,Li_PhysRevB.94.035107,Hayami_PhysRevLett.121.137202}. 
Their magnitudes and signs depend on the detailed electronic band structures.  
Hereafter, we mainly focus on the easy-axis anisotropic case with $I^z=0.2$ unless otherwise noted, since it is well known that the easy-axis anisotropy favors the SkX in centrosymmetric magnets~\cite{leonov2015multiply,Lin_PhysRevB.93.064430,Hayami_PhysRevB.93.184413,Hayami_PhysRevB.99.094420,hayami2020multiple}. 
We also focus on the case with $I^{\rm BA}>0$, since qualitatively similar results are obtained for $I^{\rm BA}<0$ by exchanging the $x$ and $y$ spin components. 

The magnetic phase diagram of the model in Eq.~(\ref{eq:Model}) is obtained for the system size with $N=96^2$ by carrying out simulated annealing in the following procedures. 
First, we start from a random spin configuration from high temperature $T_0=1.0$-$10.0$. 
Then, we reduce the temperature with the rate $T_{n+1}=\alpha T_{n}$, where $T_n$ is the temperature in the $n$th step and $\alpha=0.99995$-$0.99999$.
At each temperature, we perform the standard Metropolis local updates in real space.
The final temperature, which is typically taken at $T=0.01$, is reached by spending totally $10^5$-$10^6$ Monte Carlo sweeps. 
Finally, we perform $10^5$-$10^6$ Monte Carlo sweeps for measurements at the final temperature, after $10^5$-$10^6$ steps for thermalization. 
We also start the simulations from the spin patterns obtained at low temperatures to determine the phase boundaries between different magnetic states.

\begin{figure}[t!]
\begin{center}
\includegraphics[width=1.0 \hsize]{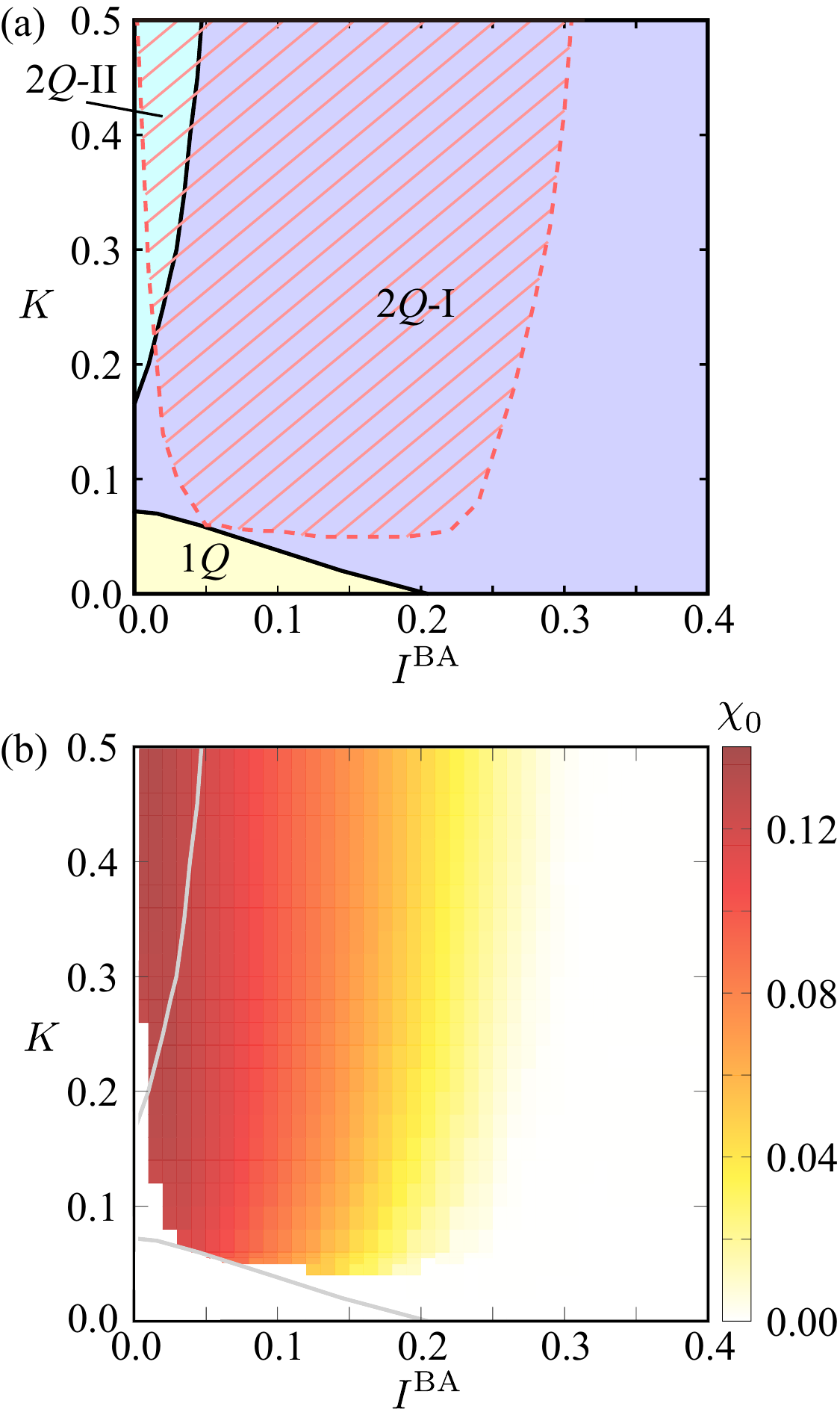} 
\caption{
\label{Fig:PD}
(a) Magnetic phase diagram at zero magnetic field for $I^{z}=0.2$ obtained by the simulated annealing down to $T=0.01$. 
2$Q$-I and 2$Q$-II stand for two different double-$Q$ states, while $1Q$ is for the single-$Q$ state. 
The hatched area shows the parameter region where the system undergoes a phase transition to a double-$Q$ state with nonzero scalar chirality in an applied magnetic field, which is deduced to realize the SkX in the ground state. 
(b) Contour plot of the maximum value of $\chi_0$ while varying $H$. 
The gray lines are the phase boundaries in (a).
}
\end{center}
\end{figure}

\begin{figure*}[htb!]
\begin{center}
\includegraphics[width=1.0 \hsize]{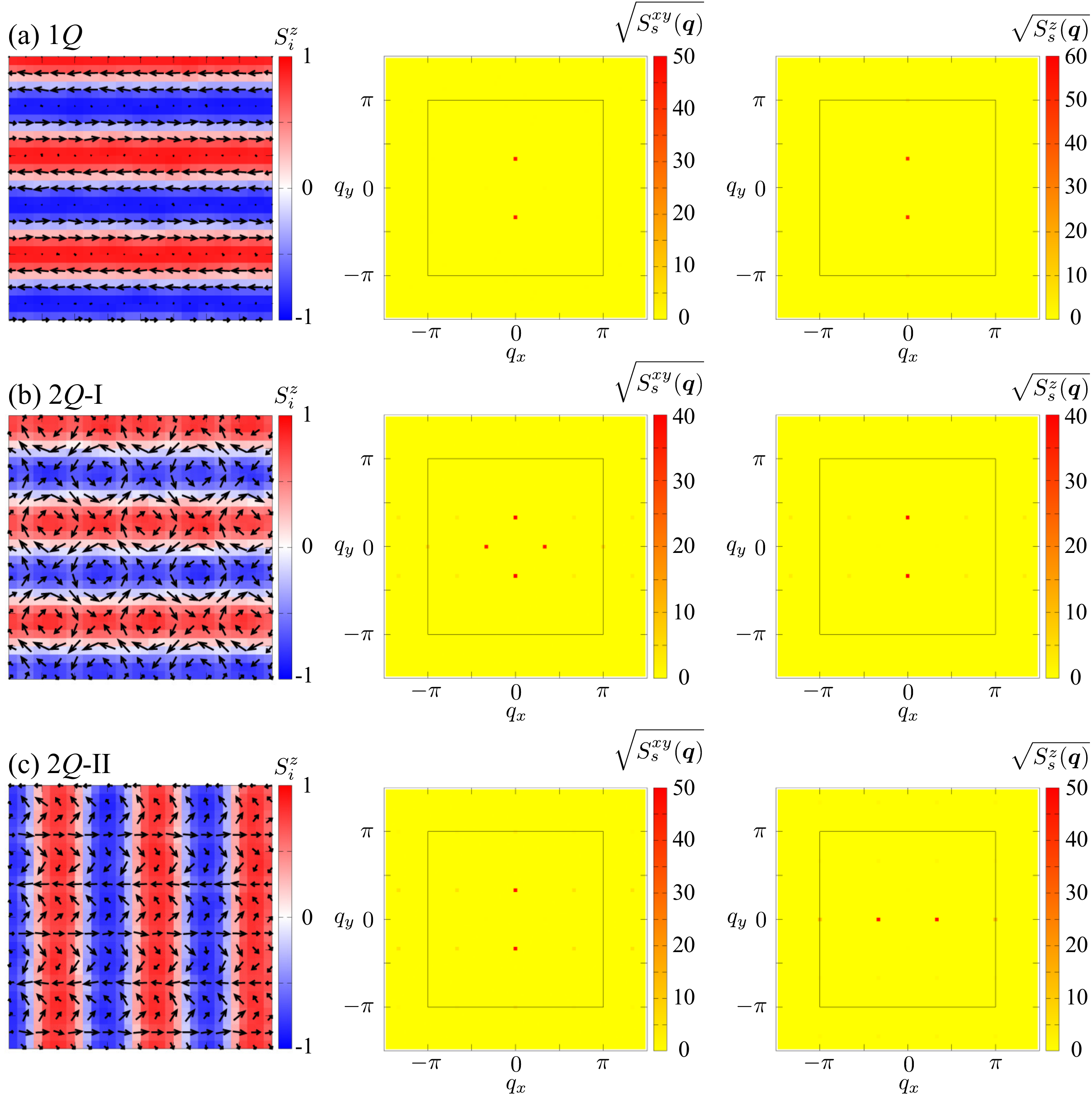} 
\caption{
\label{Fig:spin}
(Left) Snapshots of the spin configurations in (a) the 1$Q$ state for $K=0.025$ and $I^{\rm BA}=0.1$, (b) the 2$Q$-I state for $K=0.15$ and $I^{\rm BA}=0.2$, and (c) the 2$Q$-II state for $K=0.5$ and $I^{\rm BA}=0.04$. 
The arrows and the contour show the $xy$ and $z$ components of the spin moment, respectively. 
(Middle and right) 
The square root of the $xy$ and $z$ components of the spin structure factor, respectively. 
The black solid squares represent the first Brillouin zone. 
}
\end{center}
\end{figure*}

In order to identify each magnetic phase, we examine the spin and chirality configurations in the obtained states. 
The spin structure factor is given by 
\begin{align}
S_s^{\alpha}(\bm{q})&= \frac{1}{N} \sum_{j,l} S_j^{\alpha} S_l^{\alpha}  e^{i \bm{q}\cdot (\bm{r}_j-\bm{r}_l)},
\end{align}
with $\alpha=x,y,z$. 
For the in-plane component, we use the notation 
\begin{align}
S_s^{xy}(\bm{q})= S_s^{x}(\bm{q})+S_s^{y}(\bm{q}). 
\end{align}
We also introduce the magnetic moments at $\bm{q}$ component as 
\begin{equation}
m^{\alpha}_{\bm{q}}=\sqrt{\frac{S^{\alpha}_s(\bm{q})}
{N}}. 
\end{equation}
In order to distinguish the in-plane components parallel and perpendicular to $\bm{Q}_1$ and $\bm{Q}_2$, we use local coordinate frames for the $\bm{Q}_1$ and $\bm{Q}_2$ components as 
\begin{equation}
\bm{m}_{\bm{Q}_{\eta}}=(m^{\parallel}_{\bm{Q}_{\eta}}, m^{\perp}_{\bm{Q}_{\eta}},m^{z}_{\bm{Q}_{\eta}}), 
\end{equation}
for $\eta=1$ and $2$, where $m^{\parallel}_{\bm{Q}_{\eta}}$ and $m^{\perp}_{\bm{Q}_{\eta}}$ are the in-plane parallel and perpendicular components, respectively. 
We also compute the net magnetization along the $z$ direction 
\begin{equation}
m_0=
\frac{1}{N}
\sum_i S_i^z. 
\end{equation}
On the other hand, the scalar chirality $\chi_0$ is evaluated by 
\begin{equation}
\chi_0=\bigg[ \frac{1}{N}
\sum_{i,\delta=\pm1}\bm{S}_i \cdot (\bm{S}_{i+\delta \hat{x}}\times \bm{S}_{j+\delta \hat{y}})\bigg]^2, 
\end{equation}
where 
$\hat{x}$ ($\hat{y}$) is the unit vector in the $x$ ($y$) direction~\cite{Yi_PhysRevB.80.054416}.

\section{Zero-field phase diagram}
\label{sec:Zero-field phase diagram}

First, we discuss the result in the absence of the magnetic field, $H=0$. 
Figure~\ref{Fig:PD}(a) shows the magnetic phase diagram while varying $I^{\rm BA}$ and $K$ at $I^z=0.2$ obtained by the simulated annealing down to $T=0.01$. 
There are three magnetic phases, whose spin configurations in real space and the spin structure factors in momentum space are shown in Fig.~\ref{Fig:spin}. 
Note that each state is energetically degenerate with the one obtained by $90^\circ$ degree rotation in the $xy$ plane because of the fourfold rotational symmetry of the system. 
The three magnetic states do not have a net scalar chirality $\chi_0$.

In the region for small $I^{\rm BA}$ and $K$, the single-$Q$ (1$Q$) state is stabilized. 
At $I^{\rm BA}=0$, the 1$Q$ state is characterized by an elliptical spiral in either the $xz$ or $yz$ plane. 
Reflecting the easy-axis anisotropy by $I^z$, the $z$ component of the spin structure factor is larger than the $xy$ component. 
A nonzero $I^{\rm BA}$ sets the spiral plane perpendicular to the ordering vector, i.e., the $xz$ ($yz$) plane for the ordering vector $\bm{Q}_2$ ($\bm{Q}_1$); 
the state with $\bm{Q}_2$ is shown in Fig.~\ref{Fig:spin}(a). 
Thus, the $1Q$ state has an elliptical proper-screw spiral. 

For larger $I^{\rm BA}$ and $K$, two types of the double-$Q$ (2$Q$) state are realized. 
The 2$Q$-I state occupies the largest portion of the phase diagram, adjacent to the 1$Q$ state upon increasing $I^{\rm BA}$ and $K$ in Fig.~\ref{Fig:PD}(a). 
The $xy$ spin component is characterized by the double-$Q$ peaks with different intensities, while the $z$ spin component is characterized by the single-$Q$ peak, as shown in the right two panels in Fig.~\ref{Fig:spin}(b). 
The real-space spin configuration in the left panel in Fig.~\ref{Fig:spin}(b) indicates that the spin texture in the 2$Q$-I state is represented by a superposition of the proper-screw spiral along the $\bm{Q}_2$ direction and the sinusoidal wave along the $\bm{Q}_1$ direction. 
The double-$Q$ structure in the $xy$ spin component leads to a periodic array of vortices. 
Although this state does not have a net scalar chirality, it exhibits the chirality density wave along the $\bm{Q}_1$ direction~\cite{Solenov_PhysRevLett.108.096403,Ozawa_doi:10.7566/JPSJ.85.103703,yambe2020double}. 

For large $K$ and small $I^{\rm BA}$, the other double-$Q$ state denoted as $2Q$-II appears in the phase diagram in Fig.~\ref{Fig:PD}(a). 
In this state, both $xy$ and $z$ components of the spin structure factor exhibit the single-$Q$ peak at $\bm{Q}_2$ and $\bm{Q}_1$, respectively, as shown in Fig.~\ref{Fig:spin}(c). 
From the real-space spin structure, the spin pattern is represented by a superposition of the sinusoidal wave along the $\bm{Q}_1$ direction in the $z$-spin component and the 
cycloidal spiral along the $\bm{Q}_2$ direction in the $xy$-spin component. 
This state also exhibits the chirality density wave along the $\bm{Q}_1$ direction~\cite{Solenov_PhysRevLett.108.096403,Ozawa_doi:10.7566/JPSJ.85.103703,yambe2020double}.

\section{Skyrmion crystal in a field}
\label{sec:Skyrmion crystal in a finite field}

\begin{figure}[htb!]
\begin{center}
\includegraphics[width=1.0 \hsize]{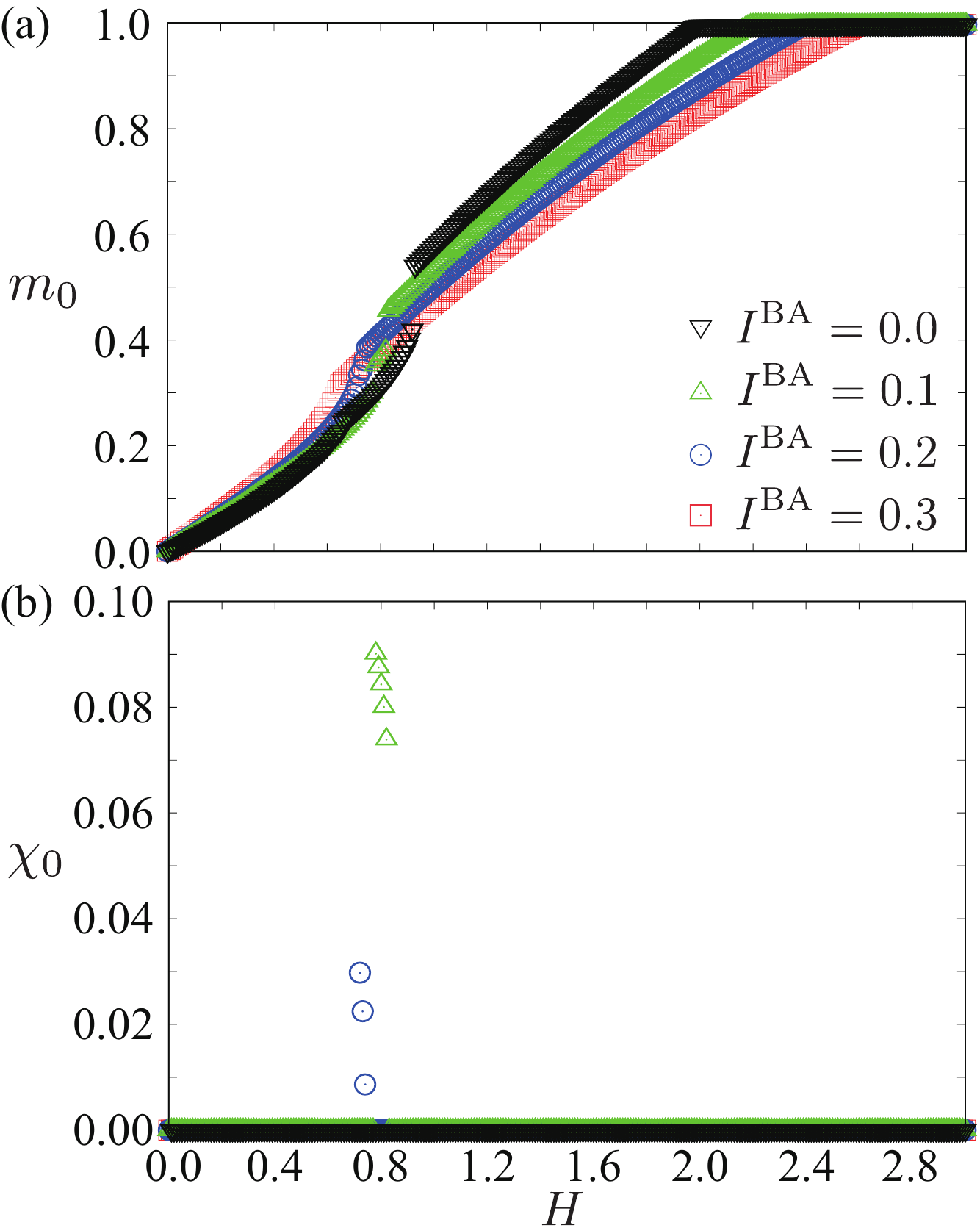} 
\caption{
\label{Fig:Mag_H_bondchange}
$H$ dependence of (a) $m_0$ and (b) $\chi_0$ for $I^{\rm BA}=0$, $0.1$, $0.2$, and $0.3$ at $K=0.2$ and $I^z=0.2$. 
}
\end{center}
\end{figure}

\begin{figure}[htb!]
\begin{center}
\includegraphics[width=0.95 \hsize]{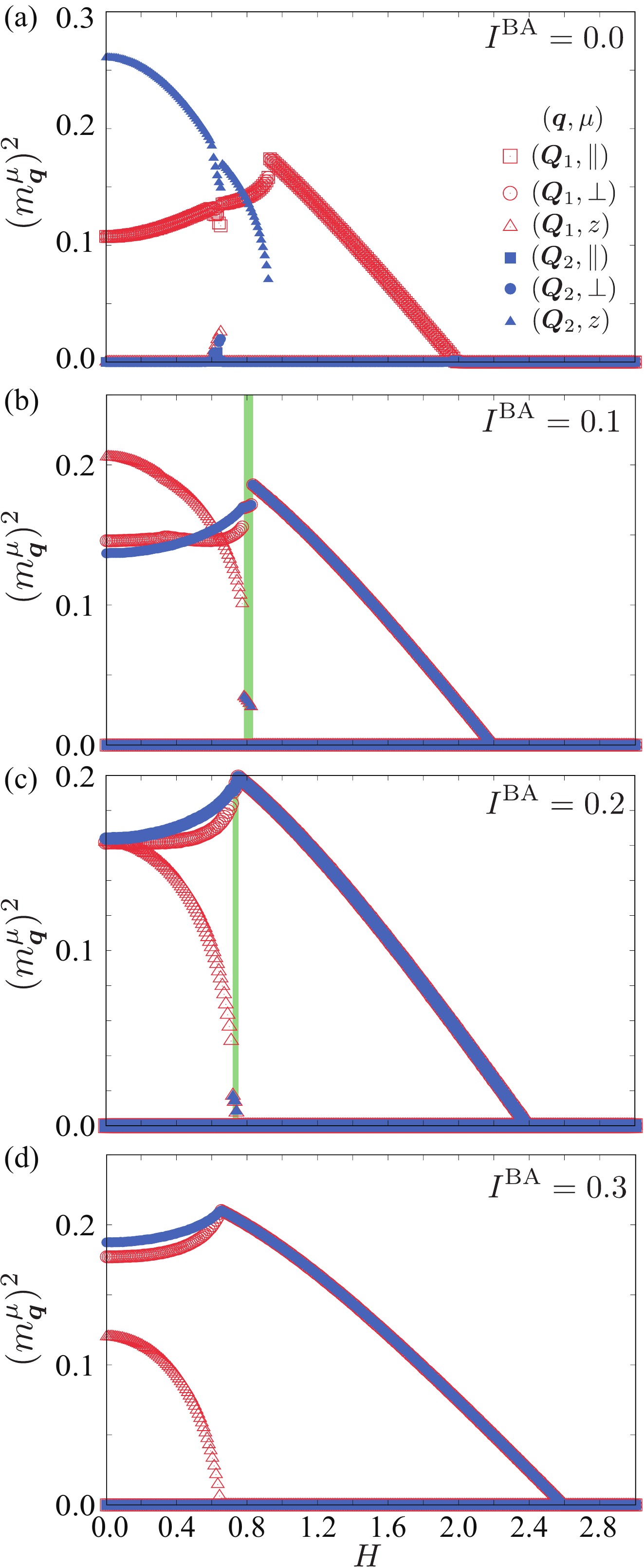} 
\caption{
\label{Fig:Mq_H_bondchange}
(a)-(d) $(m^{\mu}_{\bm{q}})^2$ ($\mu=\parallel, \perp, z$ and $\bm{q}=\bm{Q}_1, \bm{Q}_2$) for (a) $I^{\rm BA}=0$, (b) $I^{\rm BA}=0.1$, (c) $I^{\rm BA}=0.2$, and (d) $I^{\rm BA}=0.3$ at $K=0.2$ and $I^z=0.2$.
The green regions in (b) and (c) indicate the states with nonzero $\chi_0$. 
}
\end{center}
\end{figure}

\begin{figure*}[htb!]
\begin{center}
\includegraphics[width=0.85 \hsize]{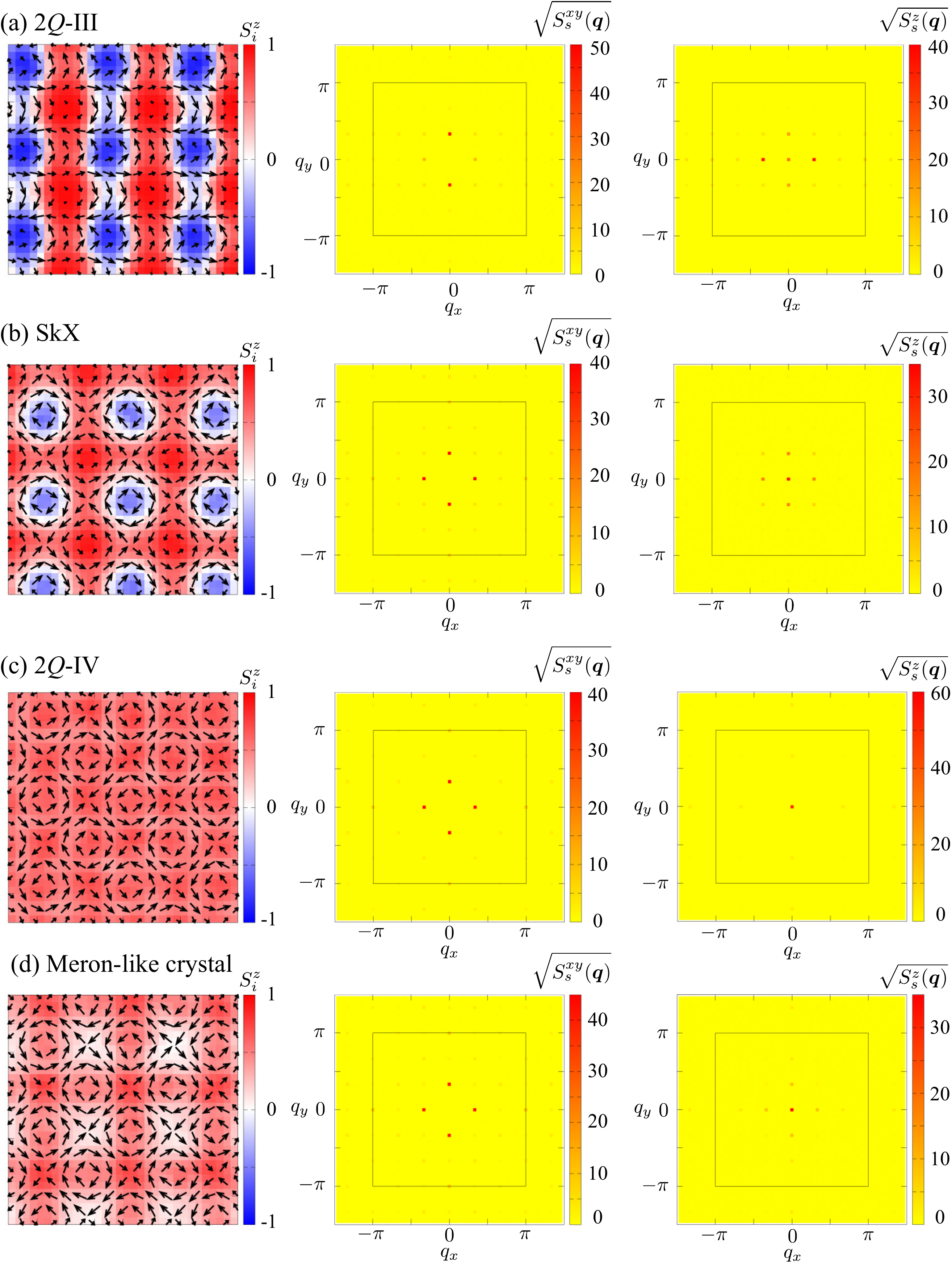} 
\caption{
\label{Fig:spin_H_bondchange}
(Left) Snapshots of the spin configurations in (a) the 2$Q$-III state for $I^{\rm BA}=0$ and $H=0.65$, (b) the SkX for $I^{\rm BA}=0.1$ and $H=0.78$, (c) the 2$Q$-IV state for $I^{\rm BA}=0.1$ and $H=1$, and (d) the meron-like crystal for $I^{\rm BA}=0.2$ and $H=0.74$ at $K=0.2$. 
The arrows and the contour show the $xy$ and $z$ components of the spin moment, respectively. 
(Middle and right) 
The square root of the $xy$ and $z$ components of the spin structure factor, respectively. 
The black solid squares represent the first Brillouin zone. 
}
\end{center}
\end{figure*}

Next, we discuss the result in the presence of the magnetic field $H$. 
From the results obtained by the simulated annealing down to $T=0.01$, we find that the system undergoes a phase transition to a double-$Q$ state with nonzero scalar chirality $\chi_0$ under the magnetic field 
in the hatched area in Fig.~\ref{Fig:PD}(a). 
The maximum value of $\chi_0$ in the field is plotted in Fig.~\ref{Fig:PD}(b). 
As detailed later, the field-induced double-$Q$ state is deduced to realize a square-type SkX in the ground state. 
The region spans both $2Q$-I and $2Q$-II states; we could not find the instability toward the SkX in the $1Q$ region. 

In the following, we discuss the detailed changes of the spin textures for the magnetic field mainly in this region. 
Interestingly, the region is drastically extended down to the small $K$ region by introducing $I^{\rm BA}$; it is limited to $K \gtrsim 0.58$ at $I^{\rm BA}=0$ (not shown), whereas the boundary comes down to $K \simeq 0.07$ for $I^{\rm BA} \simeq 0.05$. 
This indicates the importance of the bond-dependent anisotropic interaction $I^{\rm BA}$ for the stabilization of the square SkX. 
We show the result while changing $I^{\rm BA}$ in Sec.~\ref{sec:Effect of bond-dependent anisotropic interaction}. 
Then, we discuss the effects of the biquadratic interaction $K$ in Sec.~\ref{sec:Effect of biquadratic interaction} and the easy-axis anisotropic interaction $I^z$ in Sec.~\ref{sec:Effect of easy-axis anisotropic interaction}. 
In these sections, we mainly focus on the region where the $2Q$-I state is stable at zero field, which appears to be relevant to the experiment in GdRu$_2$Si$_2$ as discussed in Sec.~\ref{sec:Relevance with GdRu$_2$Si$_2$}; the $2Q$-II region with smaller $I^{\rm BA}$ and larger $K$ is discussed in Appendix~\ref{sec:appendix}.

\subsection{Effect of bond-dependent anisotropic interaction}
\label{sec:Effect of bond-dependent anisotropic interaction}

Figures~\ref{Fig:Mag_H_bondchange} and \ref{Fig:Mq_H_bondchange} show the magnetic field dependence of the spin- and chirality-related quantities for several values of $I^{\rm BA}$ at $K=0.2$ and $I^z=0.2$. 
In the simulations, as in the case of zero field, energetically-degenerate magnetic states are obtained from different initial configurations owing to the fourfold rotational symmetry;  
e.g., the single-$Q$ state with $\bm{m}_{\bm{Q}_1} \neq 0$ is equivalent to that with $\bm{m}_{\bm{Q}_2} \neq 0$. 
For better readability, we show the spin texture in each ordered state by appropriately sorting $(\bm{m}_{\bm{Q}_\nu})^2$ in Fig.~\ref{Fig:Mq_H_bondchange} and hereafter. 

At $I^{\rm BA}=0$, where the 2$Q$-II state is stabilized at $H=0$, the dominant $(m^{z}_{\bm{Q}_2})^2$ is suppressed and the subdominant $(m^{\parallel}_{\bm{Q}_1})^2$ and $(m^{\perp}_{\bm{Q}_1})^2$ are enhanced while increasing $H$, as shown in Fig.~\ref{Fig:Mq_H_bondchange}(a). 
In the narrow range of $0.60 \lesssim H \lesssim 0.65$, a different double-$Q$ (2$Q$-III) state is stabilized, whose real-space spin configuration and spin structure factor are shown in Fig.~\ref{Fig:spin_H_bondchange}(a). 
Compared to the 2$Q$-II state, the 2$Q$-III state has additional magnetic moments in $(m^{z}_{\bm{Q}_1})^2$, $(m^{\parallel}_{\bm{Q}_2})^2$, and $(m^{\perp}_{\bm{Q}_2})^2$, as shown in Fig.~\ref{Fig:Mq_H_bondchange}(a). 
The net magnetization $m_0$ shows a small anomaly corresponding to the appearance of the $2Q$-III state, as shown in Fig.~\ref{Fig:Mag_H_bondchange}(a). 
Upon further increasing $H$, the 2$Q$-II state appears again for $H \gtrsim 0.65$, which turns into the single-$Q$ conical spiral state at $H\simeq 0.93$ with a jump of $m_0$, as shown in Figs.~\ref{Fig:Mag_H_bondchange}(a) and \ref{Fig:Mq_H_bondchange}(a). 
The single-$Q$ conical state continuously changes into the fully-polarized state at $H \simeq 2$.
Throughout all these spin states, $\chi_0$ is always zero, as shown in Fig.~\ref{Fig:Mag_H_bondchange}(b) (see also Fig.~\ref{Fig:PD}(b). 

For $I^{\rm BA}=0.1$ and $0.2$, however, we find another double-$Q$ state with nonzero $\chi_0$ in a magnetic field. 
In both cases, we obtain three double-$Q$ states in addition to the fully-polarized state, as shown in Figs.~\ref{Fig:Mq_H_bondchange}(b) and \ref{Fig:Mq_H_bondchange}(c). 
The low-field state corresponds to the 2$Q$-I state connected to that at $H=0$ 
[see Fig.~\ref{Fig:spin}(b)], while the high-field state before entering the fully-polarized state corresponds to a different double-$Q$ (2$Q$-IV) state, whose spin structure is shown in Fig.~\ref{Fig:spin_H_bondchange}(c). 
This 2$Q$-IV state exhibits the double-$Q$ peaks at $(m^{\perp}_{\bm{Q}_1})^2$ and $(m^{\perp}_{\bm{Q}_2})^2$ in addition to the uniform magnetization. 
In other words, this state is characterized by a superposition of two sinusoidal waves along the $\bm{Q}_1$ and $\bm{Q}_2$ directions. 
We note that a similar spin texture was also obtained even without $I^{\rm BA}$ by considering large $K$~\cite{Hayami_PhysRevB.95.224424}.

The intermediate-field state, which is sandwiched by the 2$Q$-I and 2$Q$-IV states, shows nonzero $\chi_0$, as shown in Fig.~\ref{Fig:Mag_H_bondchange}(b).  
The phase transitions between these three double-$Q$ states are of first order with discontinuities in $\chi_0$ as well as $m_0$. 
The spin structure of the intermediate state in the case of $I^{\rm BA}=0.1$ is shown in Fig.~\ref{Fig:spin_H_bondchange}(b). 
It is a square-type SkX with fourfold rotational symmetry, composed of the equal weights for $\bm{Q}_1$ and $\bm{Q}_2$ in both $xy$- and $z$-spin components. 
Indeed, we find that the skyrmion number for this state asymptotically approaches $\pm 1$ while lowering temperature (not shown). 
Note that the SkX is energetically degenerate with the anti-skyrmion counterpart in the present model; 
the degeneracy can be lifted by including contributions from higher harmonics, as discussed in Ref.~\onlinecite{Hayami_doi:10.7566/JPSJ.89.103702}. 

The results are overall similar for $I^{\rm BA}=0.2$, as shown in Figs.~\ref{Fig:Mag_H_bondchange} and \ref{Fig:Mq_H_bondchange}(c). 
We note, however, that the field range of the intermediate double-$Q$ state becomes narrow and $\chi_0$ is reduced compared with those for $I^{\rm BA}=0.1$, since $I^{\rm BA}$ tends to forces the spins to lie in a plane; 
actually, the skyrmion number obtained at $T=0.01$ decreases while increasing $I^{\rm BA}$ in the hatched region in Fig.~\ref{Fig:PD}(a) (not shown). 
In the present simulation for $I^{\rm BA}=0.2$, the intermediate state with nonzero $\chi_0$ has the absolute value of the skyrmion number close to 1 in the region close to the phase boundary with the lower-field 2$Q$-I state, but it is reduced to less than $0.5$ when approaching the phase boundary with the higher-field 2$Q$-IV state. 
Interestingly, the spin configuration with the reduced skyrmion number less than 0.5 is characterized by the meron-crystal-like one as shown in Fig.~\ref{Fig:spin_H_bondchange}(d), which has a periodic swirling spin texture as the SkX but all the spins have positive $z$-spin moments~\cite{Lin_PhysRevB.91.224407,yu2018transformation,Borge_PhysRevB.99.060407,Bera_PhysRevResearch.1.033109,hayami2020multiple}.

The results with non-quantized skyrmion number indicate that the temperature in our simulated annealing is not sufficiently low to reach the ground state. 
From the temperature dependence of the skyrmion number, however, we conclude that the system exhibits the square-type SkX with the quantized skyrmion number of $\pm 1$ in most of the hatched region in Fig.~\ref{Fig:PD}(a) except for a narrow window with large $I^{\rm BA}$. 
For instance, the window ranges for $0.25 \lesssim I^{\rm BA} \lesssim 0.27$ at $K=0.2$. 
In the narrow window, there are, at least, two possibilities inferred from the fact that the double-$Q$ state can take not only the SkX but also the meron crystal with skyrmion number of $\pm 1/2$ depending on the way of superposition of the $\bm{Q}_1$ and $\bm{Q}_2$ helices~\cite{berg1981definition}. 
One is that we reach the SkX at the lowest temperature in all the hatched area including the narrow range. 
The other is that the ground state in the narrow range (or a part of it) is not the SkX but the meron crystal. 
In the latter case, we may have a phase transition between the SKX and meron crystal by changing the magnetic field. 
To clarify this subtle issue, we need further studies at lower temperature, which are computationally laborious.

When increasing $I^{\rm BA}$ outside the hatched region in Fig.~\ref{Fig:PD}(a), the intermediate state with nonzero $\chi_0$ vanishes, as exemplified for $I^{\rm BA}=0.3$ in Figs.~\ref{Fig:Mag_H_bondchange} and \ref{Fig:Mq_H_bondchange}(d). 
In this case, the $2Q$-I state continuously changes into the 2$Q$-IV state.

\begin{figure}[htb!]
\begin{center}
\includegraphics[width=1.0 \hsize]{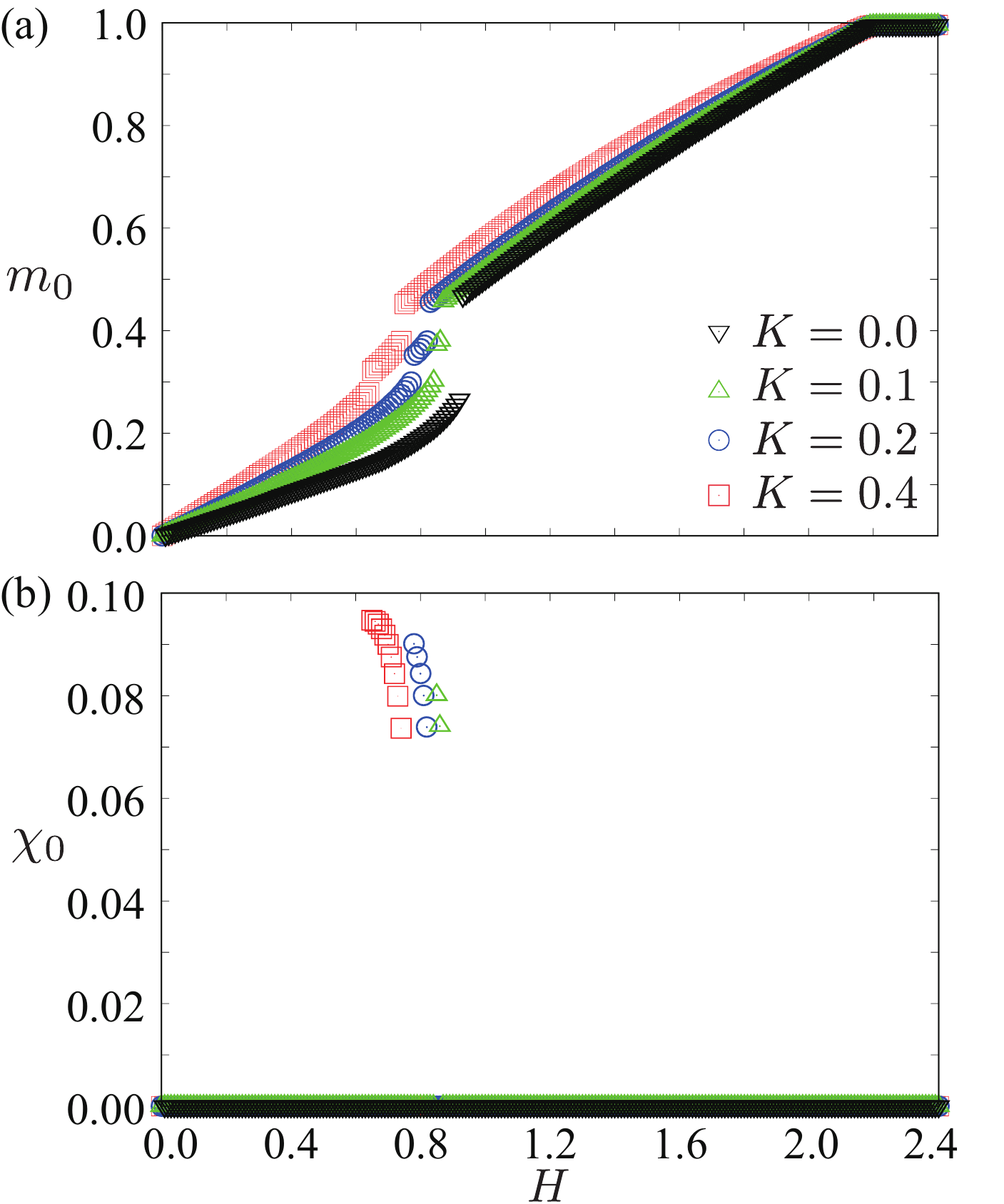} 
\caption{
\label{Fig:Mag_H_Kchange}
$H$ dependence of (a) $m_0$ and (b) $\chi_0$ for $K=0$, $0.1$, $0.2$, and $0.4$ at $I^{\rm BA}=0.1$ and $I^z=0.2$. 
}
\end{center}
\end{figure}

\begin{figure}[htb!]
\begin{center}
\includegraphics[width=1.0 \hsize]{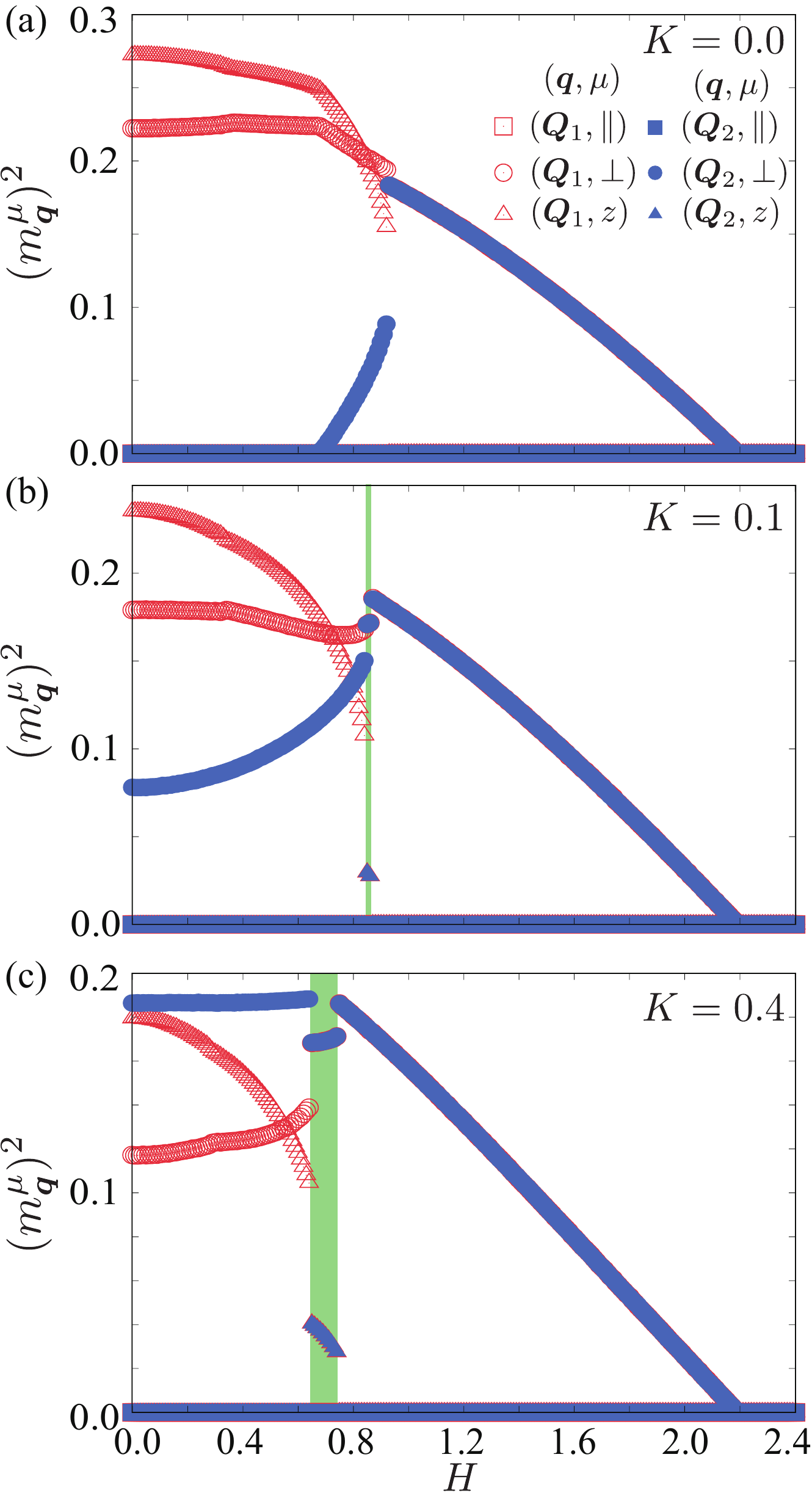} 
\caption{
\label{Fig:Mq_H_Kchange}
(a)-(c) $(m^{\mu}_{\bm{q}})^2$ ($\mu=\parallel, \perp, z$ and $\bm{q}=\bm{Q}_1, \bm{Q}_2$) for (a) $K=0$, (b) $K=0.1$, and (c) $K=0.4$ at $I^{\rm BA}=0.1$ and $I^z=0.2$.
The result at $K=0.2$ is shown in Fig.~\ref{Fig:Mq_H_bondchange}(b). 
The green regions in (b) and (c) indicate the states with nonzero $\chi_0$. 
}
\end{center}
\end{figure}

\subsection{Effect of biquadratic interaction}
\label{sec:Effect of biquadratic interaction}

Next, we discuss the behavior while changing $K$. 
Figures~\ref{Fig:Mag_H_Kchange} and \ref{Fig:Mq_H_Kchange} show the magnetic field dependence of the spin- and chirality-related quantities for $K=0$, $0.1$, $0.2$, and $0.4$ at $I^{\rm BA}=0.1$ and $I^z=0.2$. 
At $K=0$, the 1$Q$ state is stabilized at $H=0$, as shown in Fig.~\ref{Fig:PD}(a). 
While increasing $H$, the 1$Q$ state continuously turns into the 2$Q$-I state at $H\simeq 0.68$, and then, there is a first-order phase transition to the 2$Q$-IV state at $H \simeq 0.93$, as shown in Fig.~\ref{Fig:Mq_H_Kchange}(a). 
The 2$Q$-IV state changes into the fully-polarized state at $H\simeq 2.2$. 
$m_0$ shows a jump at the transition from $2Q$-I to $2Q$-IV, as shown in Fig.~\ref{Fig:Mag_H_Kchange}(a). 
$\chi_0$ is always zero as shown in Fig.~\ref{Fig:Mag_H_Kchange}(b). 

Meanwhile, for $K=0.1$, $0.2$, and $0.4$, where the $2Q$-I state is stabilized at zero field as shown in Fig.~\ref{Fig:PD}(a), the square SkX phase appears in the intermediate-field region. 
The phase sequence while increasing $H$ is similar to those in Sec.~\ref{sec:Effect of bond-dependent anisotropic interaction}, namely, from $2Q$-I, SkX, $2Q$-IV, and finally to the fully-polarized state, as shown 
in Fig.~\ref{Fig:Mq_H_Kchange}(b) for $K=0.1$, Fig.~\ref{Fig:Mq_H_bondchange}(b) for $K=0.2$, and Fig.~\ref{Fig:Mq_H_Kchange}(c) for $K=0.4$. 
The emergence of the SkX is signaled by nonzero $\chi_0$ in Fig.~\ref{Fig:Mag_H_Kchange}(b) as well as the jumps in $m_0$ in Fig.~\ref{Fig:Mag_H_Kchange}(a). 
The maximum value of $\chi_0$ becomes larger for larger $K$, as shown in Fig.~\ref{Fig:Mag_H_Kchange}(b). 
At the same time, the field range of the SkX state also becomes wider for larger $K$. 
These indicate that the biquadratic interaction $K$ originating from the itinerant nature of electrons plays an important role in the stabilization of the SkX, as in the previous studies~\cite{Hayami_PhysRevB.95.224424,Ozawa_PhysRevLett.118.147205,hayami2020multiple,Okumura_PhysRevB.101.144416}.

\begin{figure}[htb!]
\begin{center}
\includegraphics[width=1.0 \hsize]{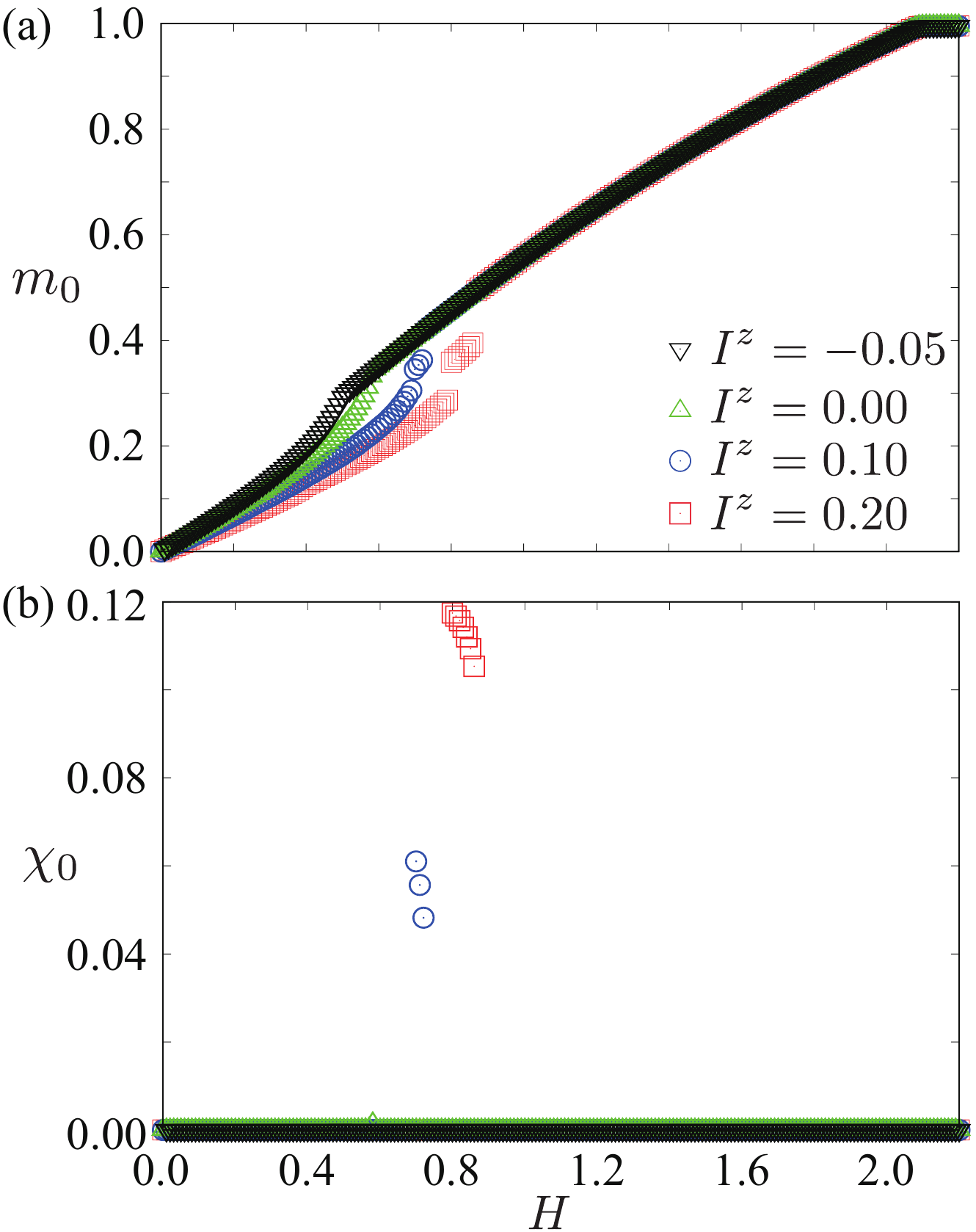} 
\caption{
\label{Fig:Mag_H_Izchange}
$H$ dependence of (a) $m_0$ and (b) $\chi_0$ for $I^z=-0.05$, $0$, $0.1$, and $0.2$ at $I^{\rm BA}=0.05$ and $K=0.2$. 
}
\end{center}
\end{figure}

\begin{figure}[htb!]
\begin{center}
\includegraphics[width=0.95 \hsize]{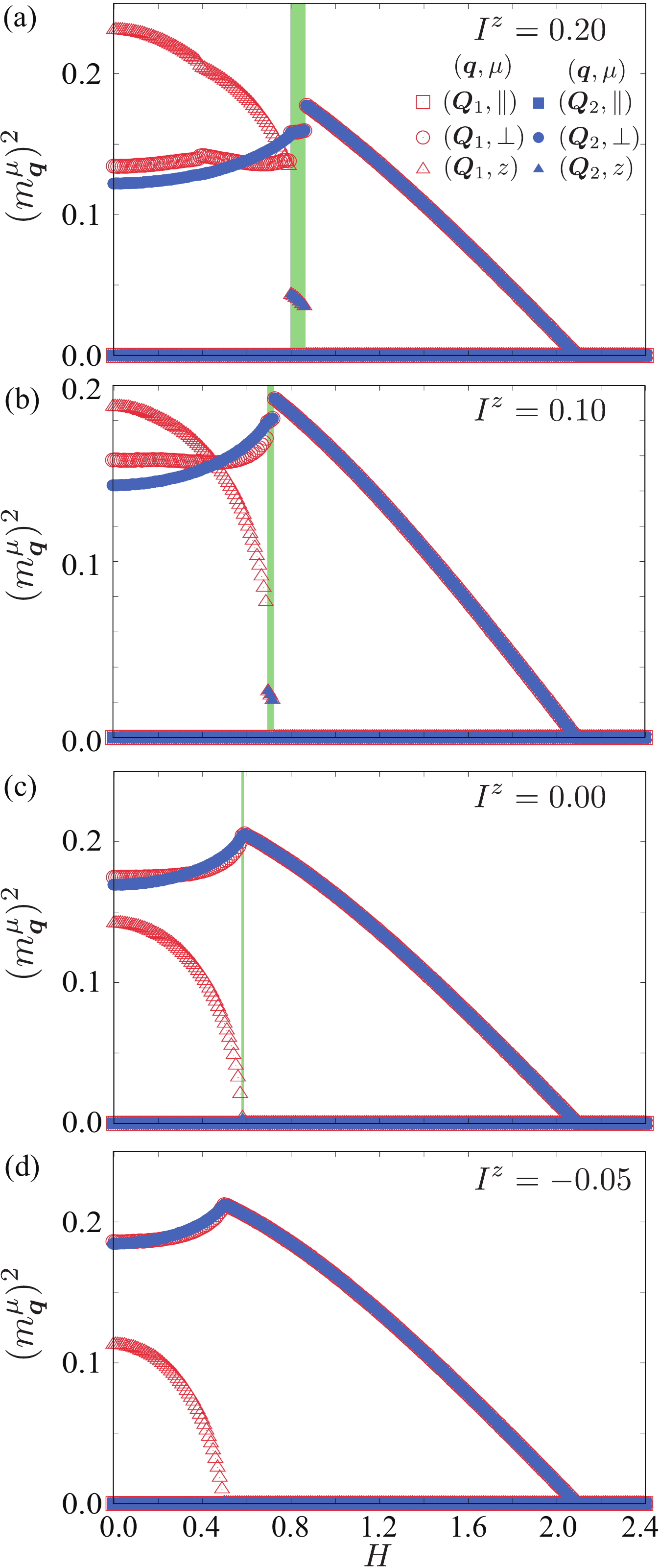} 
\caption{
\label{Fig:Mq_H_Izchange}
(a)-(d) $(m^{\mu}_{\bm{q}})^2$ ($\mu=\parallel, \perp, z$ and $\bm{q}=\bm{Q}_1, \bm{Q}_2$) for (a) $I^z=0.2$, (b) $I^z=0.1$, (c) $I^z=0$, and (d) $I^z=-0.05$ at $I^{\rm BA}=0.05$ and $K=0.2$.
The green regions in (a), (b), and (c) indicate the states with nonzero $\chi_0$. 
}
\end{center}
\end{figure}

\subsection{Effect of easy-axis anisotropic interaction}
\label{sec:Effect of easy-axis anisotropic interaction}

Lastly, we investigate the effect of $I^z$ on the SkX by considering the parameter region where the SkX is relatively robust, i.e., in the small $I^{\rm BA}$ region. 
We show the results at $I^{\rm BA}=0.05$ and $K=0.2$ while decreasing $I^{z}$ from $0.2$ to $0$ in Figs.~\ref{Fig:Mag_H_Izchange} and \ref{Fig:Mq_H_Izchange}. 
While decreasing $I^z$, the region for the SkX becomes narrower. 
For $I^z=0$, $\chi_0$ retains a tiny nonzero value only at $H \simeq 0.58$, as shown in Figs.~\ref{Fig:Mag_H_Izchange}(b). 
In the present simulation at $T=0.01$, this state exhibits the skyrmion number less than 0.5, whose spin texture is similar to that in the meron-like crystal shown in Fig.~\ref{Fig:spin_H_bondchange}(d). 
By introducing the easy-plane anisotropic interaction with $I^z=-0.05$, the region with nonzero $\chi_0$ vanishes as shown in Fig.~\ref{Fig:Mq_H_Izchange}(d). 
The results clearly indicate that the easy-axis anisotropic interaction plays an important role in the stabilization of the SkX. 
This tendency is commonly seen in centrosymmetric systems on a triangular lattice~\cite{leonov2015multiply,Lin_PhysRevB.93.064430,Hayami_PhysRevB.93.184413,Hayami_PhysRevB.99.094420}. 

\section{Discussion}
\label{sec:Discussion}

\subsection{Comparison with experiment}
\label{sec:Relevance with GdRu$_2$Si$_2$}

Let us compare our results with the recent experiments for a centrosymmetric material GdRu$_2$Si$_2$ where the square SkX was discovered in the magnetic field~\cite{khanh2020nanometric,Yasui2020}. 
In GdRu$_2$Si$_2$, three distinct phases were observed besides the fully-polarized state at high fields, which were denoted as Phase I, II, and III from the low to high magnetic field~\cite{khanh2020nanometric,Yasui2020}. 
Phase I has an anisotropic double-$Q$ structure, while Phase II and III show isotropic double-$Q$ structures. 
Among the three, Phase II shows a large topological Hall effect, and was identified as the square SkX by the Lorentz transmission electron microscopy~\cite{khanh2020nanometric}. 
The resonant x-ray scattering and the subsequent spectroscopic-imaging scanning tunneling microscopy measurements implied that the spin textures in Phase I and III were characterized by a superposition of the modulated screw and the fan structure, respectively~\cite{khanh2020nanometric, Yasui2020}. 

Our effective spin model exhibits the square SkX in the intermediate-field region similar to Phase II in GdRu$_2$Si$_2$. 
The SkX appears in a wide parameter region of $I^{\rm BA}$ and $K$ for $I^z>0$. 
Furthermore, we obtain two different types of double-$Q$ states, the 2$Q$-I and 2$Q$-IV states, in the lower- and higher-field regions of the SkX, which possess similar features to Phase I and III in GdRu$_2$Si$_2$, respectively;  
the low-field 2$Q$-I state shows the modulated screw structure consisting of the proper-screw spiral and the sinusoidal wave as shown in Fig.~\ref{Fig:spin}(b), and the high-field 2$Q$-IV state shows the fan structure consisting of the sinusoidal waves and the uniform magnetization as shown in Fig.~\ref{Fig:spin_H_bondchange}(c). 
These results indicate good agreement between Phase I, II, and III in GdRu$_2$Si$_2$ and the 2$Q$-I, SkX, and 2$Q$-IV states in our model. 

Moreover, our model analysis explains the stability of the square SkX against the other phases semiquantitatively. 
In GdRu$_2$Si$_2$, the square SkX was observed in a narrow field range between $2.1$~T and $2.5$~T, where the saturation field is around $10$~T~\cite{khanh2020nanometric}. 
Thus, the ratio of the magnetic field range where the square SkX is stabilized to the saturation field is about $4$\%. 
On the other hand, the ratio in the present model ranges is typically a few percent of the saturation field as shown in Sec.~\ref{sec:Skyrmion crystal in a finite field}, which is consistent with the experimental value. 

From these observations, we conclude that our model describes the essential physics in the centrosymmetric skyrmion material GdRu$_2$Si$_2$. 
Our results clearly indicate that the synergy between the biquadratic interaction arising from the itinerant nature of electrons, the bond-dependent anisotropic interaction, and the easy-axis anisotropic interaction plays a central role in the skyrmion physics in this compound.

\subsection{Comparison with the triangular skyrmion crystal}
\label{sec:Comparison with the triangular skyrmion crystal}

Let us compare the stability between the square and triangular SkXs in centrosymmetric itinerant electron systems. 
The triangular SkX on a triangular lattice is stabilized by taking into account either the positive biquadratic~\cite{Hayami_PhysRevB.95.224424}, the bond-dependent anisotropic~\cite{amoroso2020spontaneous, Hayami2020}, or the easy-axis anisotropic interaction~\cite{Wang_PhysRevLett.124.207201}. 
In other words, it can be stabilized by only one of the three interactions.
In stark contrast, as shown in the present study, the interplay among the three interactions is essential to realize the square SkX on a square lattice. 
Furthermore, the square SkX on a centrosymmetric lattice system has not been reported by other mechanisms thus far, in contrast to the triangular ones being realized, e.g., by frustrated exchange interactions~\cite{Okubo_PhysRevLett.108.017206,leonov2015multiply,Lin_PhysRevB.93.064430,Hayami_PhysRevB.93.184413,batista2016frustration}. 
Thus, the present square SkX is characteristic of itinerant magnets with magnetic anisotropy, which strongly suggests that the SkX observed in GdRu$_2$Si$_2$ is generated as a consequence of such a synergetic effect.

\section{Summary}
\label{sec:Summary}

We have investigated the stability of the square SkX on a centrosymmetric tetragonal lattice in itinerant magnets. 
Our results were obtained by numerically simulated annealing for an effective spin model with the long-ranged anisotropic interactions defined in momentum space. 
We found that the square SkX is stabilized by the interplay among the positive biquadratic,  bond-dependent anisotropic, and easy-axis anisotropic interactions in an external magnetic field. 
The square SkX is a double-$Q$ state composed of two helices with equal weight, retaining the fourfold rotational symmetry of  the square lattice. 
In addition, we found several different double-$Q$ states around the SkX. 
We showed that the SkX becomes more stable for larger biquadratic interaction, smaller but nonzero 
bond-dependent anisotropic interaction, and larger easy-axis anisotropic interaction. 
Our results well reproduce the three magnetic phases including the square SkX observed in GdRu$_2$Si$_2$ in the magnetic field~\cite{khanh2020nanometric,Yasui2020}, indicating the importance of the synergetic effect between the three interactions in this material. 
Our systematic study would be a reference to further exploration of skyrmion-hosting materials in centrosymmetric itinerant magnets.

\appendix

\section{Effect of magnetic field on $2Q$-II state}
\label{sec:appendix}

\begin{figure}[htb!]
\begin{center}
\includegraphics[width=1.0 \hsize]{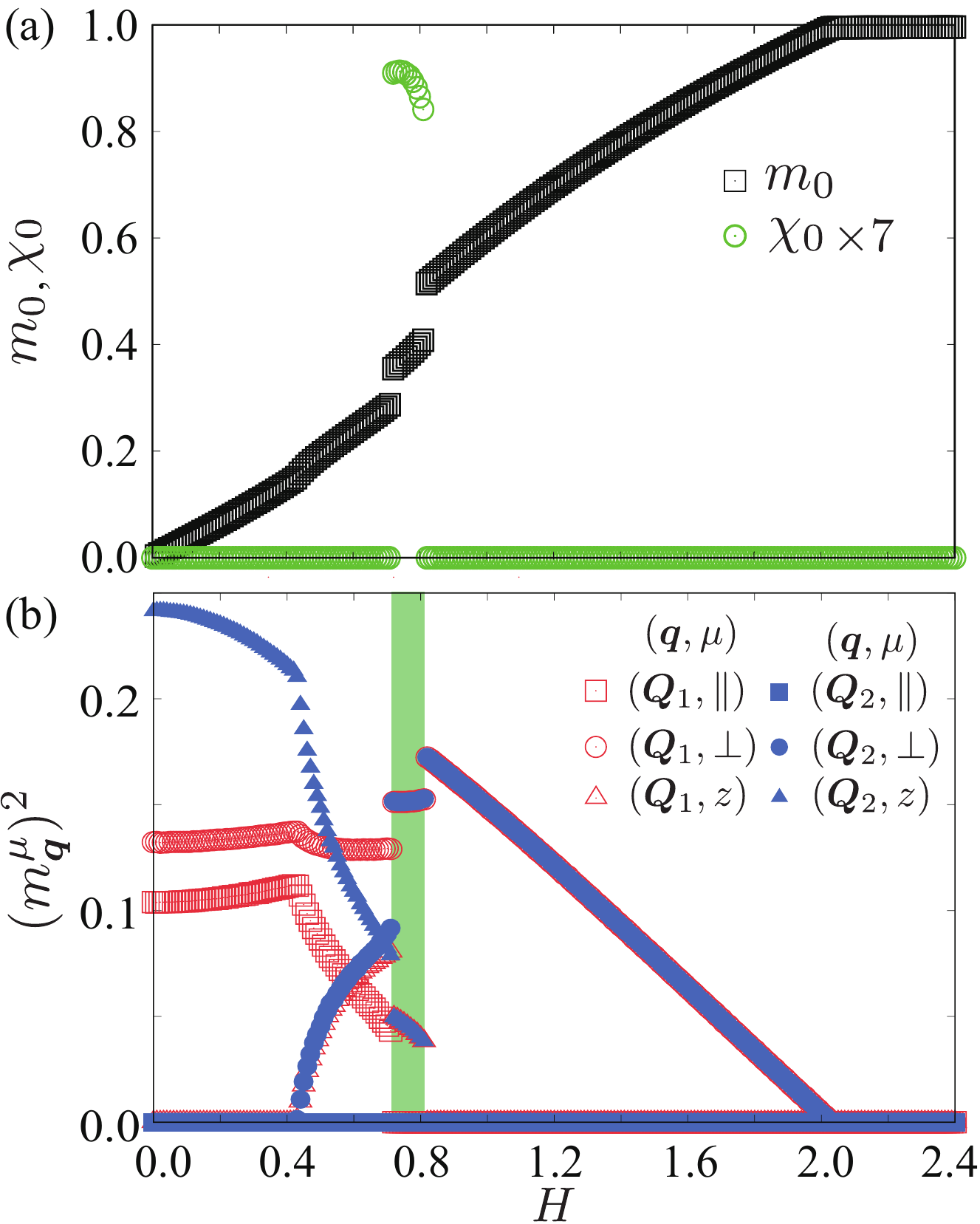} 
\caption{
\label{Fig:Mag_H_phase_2QII}
$H$ dependence of (a) $m_0$ and $\chi_0$ and 
(b) $(m^{\mu}_{\bm{q}})^2$ ($\mu=\parallel, \perp, z$ and $\bm{q}=\bm{Q}_1, \bm{Q}_2$) for $I^{\rm BA}=0.02$ at $K=0.4$ and $I^z=0.2$.
The green region in (b) indicates the states with nonzero $\chi_0$. 
}
\end{center}
\end{figure}

In this Appendix, we show the effect of the magnetic field on the $2Q$-II state within the hatched region in Fig.~\ref{Fig:PD}(a). 
We show that the square SkX is induced also in this region by the magnetic field. 
Figure~\ref{Fig:Mag_H_phase_2QII} shows the result at $I^{\rm BA}=0.02$ for $K=0.4$ and $I^z=0.2$. 
In contrasts to the result in Fig.~\ref{Fig:Mq_H_bondchange}(a) for $I^{\rm BA}=0$ and $K=0.2$, which is also the $2Q$-II state at zero field, there appear four states in addition to the fully-polarized state for $H \gtrsim 2$: the 2$Q$-II state for $0 \lesssim H\lesssim 0.44$, the 2$Q$-III state for $0.44 \lesssim H\lesssim 0.71$, the square SkX for $0.71 \lesssim H\lesssim 0.82$, and the 2$Q$-IV state for $0.82 \lesssim H\lesssim 2$, as shown in Fig.~\ref{Fig:Mag_H_phase_2QII}(b).
Their phase transitions are signaled by the kinks in $m_0$ around $H \simeq 0.44$ and $H\simeq 2$ and the jumps in $m_0$ and $\chi_0$ at $H\simeq 0.71$ and $H\simeq 0.82$, as shown in Fig.~\ref{Fig:Mag_H_phase_2QII}(a).

\begin{acknowledgments}
We thank for S. Seki, N. D. Khanh, T. Hanaguri, and Y. Yasui for fruitful discussions. 
This research was supported by JSPS KAKENHI Grants Numbers JP18K13488, JP19K03752, JP19H01834, JP19H05825, and JST CREST (JP-MJCR18T2). 
Parts of the numerical calculations were performed in the supercomputing systems in ISSP, the University of Tokyo.
\end{acknowledgments}

\bibliography{ref.bib}

\end{document}